\documentstyle[12pt]{article}

\begin{document}

\title{Partial summation of the nonlocal expansion for the
gravitational effective action in 4 dimensions}

\author{
A. G. Mirzabekian$^{a}$,
G. A. Vilkovisky$^{a,b}$,
V. V. Zhytnikov$^{a,c}$\\[3mm]
\parbox{13.5cm}{
\small $^{a}${\em Lebedev Research Center in Physics,
Leninsky Prospect 53, Moscow 117924, Russia} \\[1.5mm]
\small $^{b}${\em Lebedev Physics Institute, Academy of Sciences
of Russia, Leninsky Prospect 53, Moscow 117924, Russia} \\[1.5mm]
\small $^{c}${\em Moscow State Pedagogical University,
Davydovskii 4, Moscow 107140, Russia}
}}

\date{}

\maketitle

\def\abstractname{\ }

\begin{abstract}
\normalsize\baselineskip=6mm 
The vacuum action for the gravitational field admits a known
expansion in powers of the Ricci tensor with nonlocal operator
coefficients (form factors). We show that going over to a
different basis of curvature invariants makes possible a partial
summation of this expansion. Only the form factors of the
Weyl-tensor invariants need be calculated. The full action is
then uniquely recovered to all orders from the knowledge of the
trace anomaly. We present an explicit expression for the partially
summed action, and point out simplifications resulting in the
vertex functions. An application to the effect of the vacuum
gravitational waves is discussed.
\end{abstract}

\newpage

Our starting point is the nonlocal expansion of the effective
action for a given quantum field model in powers of a collective
field strength [1-7]. In the purely gravitational sector this field
strength boils down to the Ricci tensor $R^\mu_\nu$ since the
Riemann tensor gets expressed through $R^\mu_\nu$ via the
Bianchi identities [1-3]. For a generic field model the collective
field strength is the set
\begin{equation}
\Re = ( R^\mu_\nu,\  \hat{\cal R}_{\mu\nu},\  \hat P)
\end{equation}
which consists of the Ricci tensor, the commutator curvature
$\hat{\cal R}_{\mu\nu}=[\nabla_\mu,\nabla_\nu]$ accounting for
an arbitrary connection and satisfying the Jacobi identity,
and the potential $\hat P$ (see refs. [2-7]). Both
$\hat{\cal R}_{\mu\nu}$ and $\hat P$ are matrices in the space
of fields disturbances. At every order in $\Re$ there is a
finite basis of nonlocal curvature invariants [2], and the
effective action is their linear combination. Denoting $W$
the euclidean effective action with the term linear in the
curvature subtracted, we have [1,2]
\begin{equation}
W=\sum^\infty_{n=2}\int dx\,g^{1/2}\,\sum_i
\Gamma_i(\Box_1,\dots\Box_n,\Box_{1+2},\Box_{2+3},\dots)\,
\Re_1\dots\Re_n(i)
\end{equation}
where, for a given order $n$, $i$ labels a finite set
of basis invariants, and $\Re_1\dots\Re_n(i)$ in the
$n$th-order basis invariant number $i$. For pure gravity,
it has the structure
\begin{equation}
\Re_1\dots\Re_n(i) = M_i(\nabla_1,\dots\nabla_n)\,
R^{\scriptscriptstyle\bullet}_{\scriptscriptstyle\bullet}{}_1
\dots
R^{\scriptscriptstyle\bullet}_{\scriptscriptstyle\bullet}{}_n
\end{equation}
where
$R^{\scriptscriptstyle\bullet}_{\scriptscriptstyle\bullet}{}_k
=R^\mu_\nu(x_k)$ is the Ricci tensor at a point $x_k$,
$M_i(\nabla_1,\dots\nabla_n)$ is a monomial in covariant derivatives
whose order does not exceed $2n$, $\nabla_k$ acts on
$R^{\scriptscriptstyle\bullet}_{\scriptscriptstyle\bullet}{}_k$,
and, after the action of all operators, the points $x_k$ are made
coincident. The arguments of the form factors $\Gamma_i$ in (2),
$\Box\equiv g^{\mu\nu}\nabla_\mu\nabla_\nu$, include all
operators $\Box_k$ acting on individual Ricci tensors
$R^{\scriptscriptstyle\bullet}_{\scriptscriptstyle\bullet}{}_k$
and all operators $\Box_{k+p}$ acting on the products
$R^{\scriptscriptstyle\bullet}_{\scriptscriptstyle\bullet}{}_k
R^{\scriptscriptstyle\bullet}_{\scriptscriptstyle\bullet}{}_p$
of two Ricci tensors. For the full set of curvatures (1), the
structure of basis invariants in the expansion (2) is similar.

The basis invariants are explicitly built in [2] to third order
inclusive, and the respective form factors $\Gamma_i$ are calculated
in the one-loop approximation  for a generic model of quantum
fields [3-6]. The asymptotic behaviours of the form factors
$\Gamma_i$ at one of the arguments small and the others fixed [7]
determine the effects of coherent [8,9] and noncoherent [10,11]
vacuum radiation from the asymptotically flat systems.

When calculated by the dimensional regularization, the effective
action of conformal invariant fields produces a local trace
anomaly (for the original references and further developments see [5]).
This property can also be formulated for a generic quantum field
model with the inverse propagator of the form
\footnote{\normalsize
Here and below, $\hat1$ is the unit matrix in the
space of fields disturbances, and the sign conventions for
the curvature are $R^\mu_{\cdot\alpha\nu\beta}=\partial_\nu
\Gamma^\mu_{\alpha\beta}-\dots,$
$R_{\alpha\beta}=R^\mu_{\cdot\alpha\mu\beta},$
$R=g^{\alpha\beta}R_{\alpha\beta}$.}
\begin{equation}
g^{\mu\nu}\nabla_\mu\nabla_\nu\hat1 +
(\hat P - \frac16 R\hat1)
\end{equation}
provided that the effective action $W$ is expressed in terms of the
collective field strength (1). Assuming that, under the conformal
transformation of all fields
\begin{equation}
\delta_\sigma = \int dx\;\sigma(x)\,\left(
2g_{\mu\nu}\frac{\delta}{\delta g_{\mu\nu}}+\dots\right),
\end{equation}
the commutator curvature and potential transform as follows
\begin{equation}
\delta_\sigma \hat{\cal R}_{\mu\nu} = 0,\qquad
\delta_\sigma  \hat P = -2\sigma\hat P ,
\end{equation}
one obtains a local expression for $\delta_\sigma W$
in the universal form (eq. (28) below) [5,12].

Because the conformal transformation is inhomogeneous in the
curvature, the expansion in powers of the curvature (2) does not
preserve the exact conformal properties of the  effective action.
These properties  can only be recovered order by order by an explicit
calculation.
To third order inclusive this calculation was done for the anomaly
in two [4] and four [5] dimensions. The purpose of the present paper
is turning this shortcoming into an advantage. We shall use the
knowledge of the exact trace anomaly to perform a partial
summation of the expansion (2).

As a first step we go over to a new basis of curvature invariants
in which the gravitational field strength is represented,
instead of $R^\mu_\nu$, by the set
$C^\alpha_{\cdot\beta\gamma\sigma},R$
consisting of the Weyl
tensor and the Ricci scalar. To show that this set is complete, we
use the corollary of the Bianchi identities for the Weyl tensor
\begin{equation}
\nabla^\gamma\nabla^\alpha C_{\alpha\beta\gamma\delta} =
\frac{1}{2}\Box R_{\beta\delta} - \frac{1}{2}
\nabla_\gamma\nabla_\delta R^\gamma_\beta +
\frac{1}{12}\nabla_\beta\nabla_\delta R
-\frac{1}{12}g_{\beta\delta}\Box R
\end{equation}
to express $R^\mu_\nu$ iteratively through
$C^\alpha_{\cdot\beta\gamma\sigma}$ and $R$:
\begin{equation}
R_{\mu\nu} = C_{\mu\nu}
+\frac{1}{3}\nabla_\mu\nabla_\nu\frac{1}{\Box}R
+\frac{1}{6}g_{\mu\nu} R
+{\rm O}[\Re^2].
\end{equation}
Here
\begin{equation}
C_{\mu\nu}
\stackrel{{\rm def}}{=} \frac{2}{\Box}
\nabla^\beta\nabla^\alpha C_{\alpha\mu\beta\nu},
\end{equation}
and the relation inverse to (9) is
\footnote{\normalsize
In lorentzian spacetime, the validity of relation (10)
with $1/\Box$ replaced by the retarded Green function
$1/\Box_{\rm ret}$ [3] is limited to the in-vacuum state for
gravitons (see also below). }
\begin{eqnarray}
C_{\alpha\beta\gamma\delta} &=&
\frac{1}{\Box}\left(
\nabla_\alpha\nabla_\gamma C_{\beta\delta}
+\nabla_\beta\nabla_\delta C_{\alpha\gamma}
-\nabla_\beta\nabla_\gamma C_{\alpha\delta}
-\nabla_\alpha\nabla_\delta C_{\beta\gamma}\right)\nonumber\\
&&-\frac{1}{2}\left(
g_{\alpha\gamma} C_{\beta\delta}
+g_{\beta\delta} C_{\alpha\gamma}
-g_{\beta\gamma} C_{\alpha\delta}
-g_{\alpha\delta} C_{\beta\gamma}
\right) + {\rm O}[\Re^2].
\end{eqnarray}

Expression (8) can now be used to rewrite the basis invariants (3)
in terms of $C^\alpha_{\cdot\beta\gamma\sigma}$ and $R$.
The result is that the action (2)  becomes a sum of three
contributions
\begin{equation}
W = W_C + W_{CR} + W_R
\end{equation}
each of which is an expansion of the form (2), (3), and $W_C$
contains the basis invariants with the Weyl tensor only:
\begin{eqnarray}
W_C &=& \sum^\infty_{n=2}\int dx\,g^{1/2}\,\sum_i
\Gamma_i(\Box_1,\dots\Box_n,\Box_{1+2},\Box_{2+3},\dots)
\nonumber\\
&&\qquad\qquad\qquad\qquad\times M_i(\nabla_1,\dots \nabla_n)\,
\mbox{\bf C}_1\dots\mbox{\bf C}_n.
\end{eqnarray}
Here $\mbox{\bf C}$ is the notation for
$C^\alpha_{\cdot\beta\gamma\sigma}$ with the indices omitted,
and we do not introduce a new notation for the form factors
in (12) which are linear combinations of the form factors in (2).
The two other terms in (11) are of a similar structure and
contain the basis invariants either with both
$C^\alpha_{\cdot\beta\gamma\sigma}$ and $R$
(the term $W_{CR}$) or with $R$ only (the term $W_R$).

The main assertion of the present paper is that the effective action
$W$ is completely determined by its term $W_C$ and its trace
anomaly. When the full set of field strengths (1) is present, the
term $W_C$ should include all invariants with $\mbox{\bf C}$,
$\hat{\cal R}_{\mu\nu}$ and $\hat P$ but not R:
\begin{equation}
W_C = W_C(g_{\mu\nu},\ \mbox{\bf C},\ \hat{\cal R}_{\mu\nu},
\ \hat P).
\end{equation}
The complementary contributions to $W$ contain at least one
Ricci scalar $R$ in every term of expansion. Below we present
(i) an explicit expression  which completely recovers $W$ given
$W_C$, and (ii) a proof that this recovery is unique.
We emphasize that the term $W_C$ {\em is not conformal
invariant}.
\footnote{\normalsize
The concept of conformal invariant part of the effective action is
ill-defined. There are many different ways to decompose a given
action into conformal invariant and anomalous parts, and there
are many different actions producing one and the same trace anomaly
(see [14] and references therein).}

For accomplishing the points (i) and (ii) above, we shall use
the method of building conformal invariants proposed in [12,13].
The main tool is the solution
\begin{equation}
\phi(x|g) = 1 + \frac{1}{6}
\left(\Box - {\textstyle\frac{1}{6}}R\right)^{-1} R
\end{equation}
of the equation
\begin{equation}
\left(\Box - {\textstyle\frac{1}{6}}R\right) \phi = 0
\end{equation}
for a scalar field in an asymptotically flat euclidean spacetime
with the boundary condition $\phi=1$ at infinity. Under a conformal
transformation that becomes the identity at infinity the
solution transforms as follows
\begin{equation}
\phi(x|g_{\mu\nu} e^{2\sigma}) =
e^{-\sigma(x)} \phi(x|g_{\mu\nu}),
\end{equation}
and, therefore, the quantity
\begin{equation}
\overline{g}_{\mu\nu}(x) = \phi^2(x|g)\,g_{\mu\nu}(x)
\end{equation}
is conformal invariant. Hence, any functional of
$\overline{g}_{\mu\nu}(x)$ is conformal invariant. When used
as a metric, $\overline{g}_{\mu\nu}(x)$ possesses the property
that its Ricci scalar vanishes identically:
\begin{equation}
R(\overline{g})\equiv0.
\end{equation}
On the other hand, $\phi$ in (14) admits an expansion of the
same nature as the one used for the construction of the effective
action (2):
\begin{equation}
\phi = 1+\frac16\frac{1}{\Box}R
+\frac{1}{36}\frac{1}{\Box}\left(R\frac{1}{\Box}R\right)
+\dots\ .
\end{equation}
It is only in the framework of this expansion that the
solution (14) will be used in the present context.

We are now able to prove that if two actions of the metric,
$W'[g]$ and $W''[g]$, each in the form of expansion (2),
or (11), give rise to one and the same trace
\begin{equation}
g_{\mu\nu} \frac{\delta W'}{\delta g_{\mu\nu}} =
g_{\mu\nu} \frac{\delta W''}{\delta g_{\mu\nu}}
\end{equation}
and have one and the same form factors of the Weyl-tensor invariants
\begin{equation}
W'_C[g] = W''_C[g],
\end{equation}
then these actions coincide. Indeed, the difference
$\Delta=W'-W''$ has the form of expansions (2) and (11)
but with no pure-Weyl terms
\begin{equation}
\Delta[g] = \Delta_{CR}[g] + \Delta_R[g]
\end{equation}
and is conformal invariant. Therefore, with $\overline{g}$
defined in (17), we have
\begin{equation}
\Delta[g] = \Delta[\overline{g}]
= \Delta_{CR}[\overline{g}]
+ \Delta_{R}[\overline{g}] = 0
\end{equation}
where the vanishing takes place owing to eq. (18) and the
fact that both $\Delta_{CR}$ and $\Delta_R$ contain at least
one $R$ in every term of expansion.

There remains to be presented an explicit expression for the
full action $W$ given $W_C$. This expression is of the
following form:
\begin{eqnarray}
W &=& W_C(\phi^2g_{\mu\nu},\ \mbox{\bf C},\ \hat{\cal R}_{\mu\nu},
	  \ \phi^{-2}\hat{P})
\nonumber\\[1mm]&&
+ \frac{1}{2(4\pi)^2}\int dx\,g^{1/2}\,{\rm tr}\,
   \left\{ \frac{1}{1080} R^2 \hat1 + \frac{1}{18} R\hat{P}\right.
\nonumber\\[2mm]&&\quad
+\frac{1}{180} Z\left(C_{\alpha\beta\gamma\delta}
C^{\alpha\beta\gamma\delta}+R_{\alpha\beta}R^{\alpha\beta}
-\frac{1}{3}R^2\right)\hat1
\nonumber\\[1mm]&&\quad
+\frac{1}{180} R^{\alpha\beta}\,\nabla_\alpha Z\nabla_\beta Z\,\hat1
+\frac{1}{360}(\nabla Z)^2\left(-R+\Box Z
+\frac{1}{4}(\nabla Z)^2\right)\hat1
\nonumber\\[1mm]&&\quad
\left.+\frac{1}{12}Z\, \hat{\cal R}_{\mu\nu}\hat{\cal R}^{\mu\nu}
+\frac{1}{2}Z\, \hat{P}\hat{P}\right\}
\end{eqnarray}
where ${\rm tr}$ denotes the matrix trace,
$Z$ is the following quantity
\begin{equation}
Z= \ln\phi^2,
\end{equation}
and $\phi$ is defined in (14). The first term in (24) is built out
of the original $W_C$ in (13) by replacing the metric $g_{\mu\nu}$
{\em wherever it appears} with $\phi^2g_{\mu\nu}$, and $\hat{P}$
with $\phi^{-2}\hat P$. Specifically, since in the purely
gravitational
sector $W_C$ is given by expansion (12), the respective contribution
to (24) is of the form
\begin{eqnarray}
W_C(\phi^2g_{\mu\nu},\mbox{\bf C})
&=& \sum^\infty_{n=2}\int dx\,\overline{g}^{1/2}\,\sum_i
\Gamma_i(\overline{\vphantom{I}\Box}_1,
\dots\overline{\vphantom{I}\Box}_n,
\overline{\vphantom{I}\Box}_{1+2},
\overline{\vphantom{I}\Box}_{2+3},\dots)
\nonumber\\
&&\qquad\qquad\qquad\qquad\times M_i(\overline{\nabla}_1,\dots
\overline{\nabla}_n)\,
\mbox{\bf C}_1\dots\mbox{\bf C}_n
\end{eqnarray}
where all symbols with the bar refer to the metric
$\overline{g}_{\mu\nu}$ in (17). The explicit addition to
$W_C$ in (24) contains $Z$ up to the fourth power but, if
needed, the power can be reduced by using the equation
\begin{equation}
\Box Z + \frac{1}{2}(\nabla Z)^2 = \frac{1}{3} R
\end{equation}
that $Z$ in (25) satisfies.

To prove that expression (24) for $W$ correctly recovers
the original action, one must first show that all pure-Weyl
invariants of $W$ are contained in the original $W_C$.
This is easily seen upon the insertion of expansion (19) in
(25) and (24). Next, one must show that the action (24) gives
rise to the correct trace anomaly. Since the term with $W_C$ in
(24) is conformal invariant by construction, the anomaly is
produced by the remaining explicit terms. The direct
calculation using (6) with $\delta_\sigma$ in (5) gives
\begin{equation}
-\delta_\sigma W = \frac{1}{(4\pi)^2}\int dx\,g^{1/2}\,
\sigma(x)\,{\rm tr}\,\hat{a}_2(x,x),
\end{equation}
\begin{eqnarray}
\hat{a}_2(x,x) &=& \frac{1}{180}\Box R\,\hat1
+\frac{1}{6}\Box \hat{P}
+ \frac{1}{180}\left(
C_{\alpha\beta\gamma\delta}C^{\alpha\beta\gamma\delta}
+R_{\alpha\beta}R^{\alpha\beta}-\frac{1}{3}R^2\right)\hat1
\nonumber\\[2mm]&&\qquad
+\frac{1}{12}\hat{\cal R}_{\mu\nu}\hat{\cal R}^{\mu\nu}
+\frac{1}{2}\hat{P}\hat{P}.
\end{eqnarray}
The latter quantity is the second DeWitt coefficient [15]
at coincident points for the operator (4). This is the correct
expression for the trace anomaly in four dimensions
(see [5,12] and references therein).

The significance of the result above is in the fact that it
suffices to calculate only the form factors of the basis
invariants with the Weyl tensor (and matter field strengths);
they accumulate the dynamical information and are generally not
controlled by symmetries. The remaining form factors in the
effective action are then obtained automatically by inserting
expansion  (19) in (24). Specifically, if the form factors of
the Weyl tensor are known to third order inclusive (as is
presently the case with the one-loop effective action [3-7]),
then all contributions to $W$ containing two or three Weyl
tensors and any number of  Ricci scalars are also known.
The contributions of invariants with the Ricci scalar only
are all known in advance in the closed form. This is a partial
summation of the series (2) for the effective action.

The transition to the
$C^\alpha_{\cdot\beta\gamma\sigma}$, $R$  basis
simplifies greatly the form factors obtained in [5,6].
The one-loop effective action for the operator (4) takes the form
\begin{eqnarray}
-W &=& \frac{1}{2(4\pi)^2}\int dx\,g^{1/2}\,{\rm tr}\,
\left\{
-\frac{1}{1080}R^2\,\hat1
-\frac{1}{18}R\hat{P}
\right.\nonumber\\[2mm]&&\quad
-\frac{1}{120}
C_{\alpha\beta\gamma\delta}
\left(\ln\left(-\Box/\mu^2\right)-\frac{16}{15}\right)
C^{\alpha\beta\gamma\delta}\,\hat1
\nonumber\\[2mm]&&\quad
-\frac{1}{12}\hat{\cal R}_{\mu\nu}
\left(\ln\left(-\Box/\mu^2\right)-\frac{2}{3}\right)
\hat{\cal R}^{\mu\nu}
-\frac{1}{2}\hat{P}
\left(\ln\left(-\Box/\mu^2\right)\right)
\hat{P}
\nonumber\\[2mm]&&\quad\left.
+\sum_i\Gamma_i(-\Box_1,-\Box_2,-\Box_3)
\overline{\Re_1\Re_2\Re_3}(i)
+{\rm O}[\Re^4]\right\}
\end{eqnarray}
where the only arbitrary parameter is $\mu^2>0$,
and the new cubic invariants
$\overline{\Re_1\Re_2\Re_3}(i)$
are obtained from the old ones
${\Re_1\Re_2\Re_3}(i)$ in [5,6] by inserting the expression (8)
for $R_{\mu\nu}$. As a result, the full list of new cubic
invariants differs from the old one only by the replacement
of $R_{\mu\nu}$ with $C_{\mu\nu}$ (eq. (9)) but the form factors
of the invariants with the Ricci scalar change drastically.
In the purely gravitational sector there are five such invariants
\begin{eqnarray}
&&\overline{\Re_1\Re_2\Re_3}({9})=R_1 R_2 R_3\hat1, \quad
\overline{\Re_1\Re_2\Re_3}({11})=
   C_1^{\mu\nu}C_{2\,\mu\nu}R_3\hat1, \nonumber\\[1.5mm]
&&\overline{\Re_1\Re_2\Re_3}({22})=C_1^{\alpha\beta}
   \nabla_\alpha R_2 \nabla_\beta R_3\hat1,\quad
\overline{\Re_1\Re_2\Re_3}({23})=\nabla^\mu C_1^{\nu\alpha}
   \nabla_\nu C_{2\,\mu\alpha}R_3\hat1,\nonumber\\[1.5mm]
&&\overline{\Re_1\Re_2\Re_3}({27})=\nabla_\alpha\nabla_\beta
   C_1^{\mu\nu}\nabla_\mu\nabla_\nu C_2^{\alpha\beta} R_3\hat1
\nonumber
\end{eqnarray}
whose form factors were the most complicated ones in the table
of ref. [6]. Now they factorize into elementary functions
of two variables and take the form
\begin{eqnarray}
\Gamma_{9} &=&
\frac1{19440}
\left({2\over {\Box_3}} - {{\Box_1}\over {\Box_2\,\Box_3}}
\right)
,\\[5mm]
\Gamma_{11} &=&
\frac1{540}
\left(
{4\over {\Box_2}} - {1\over {\Box_3}}
- {{2\,\Box_1}\over {\Box_2\,\Box_3}} -
{{\Box_3}\over {\Box_1\,\Box_2}}
\right)
 + \frac1{720}
\left(
{{10\,\Box_1}\over {\Box_2}}
\right.\nonumber\\&&\quad\left.
+ {{6\,\Box_1}\over {\Box_3}}
- {{2\,{{\Box_1}^2}}\over {\Box_2\,\Box_3}} -
  {{14\,\Box_3}\over {\Box_2}} + {{3\,{{\Box_3}^2}}\over {\Box_1\,\Box_2}}
\right)
\frac{\ln(\Box_1/\Box_2)}{(\Box_1-\Box_2)}
,\\[5mm]
\Gamma_{22} &=&
{1\over {1620\,\Box_2\,\Box_3}}
,\\[5mm]
\Gamma_{23} &=&
\frac1{135}
\left(
{1\over {\Box_1\,\Box_2}} - {2\over {\Box_2\,\Box_3}}
\right)
\nonumber\\&&
 + \frac1{180}
\left(
{8\over {\Box_2}} - {{2\,\Box_1}\over {\Box_2\,\Box_3}} -
  {{3\,\Box_3}\over {\Box_1\,\Box_2}}
\right)
\frac{\ln(\Box_1/\Box_2)}{(\Box_1-\Box_2)}
,\\[5mm]
\Gamma_{27} &=&
{{-1}\over {135\,\Box_1\,\Box_2\,\Box_3}}
 + \frac1{180}
\left(
{3\over {\Box_1\,\Box_2}} - {2\over {\Box_2\,\Box_3}}
\right)
\frac{\ln(\Box_1/\Box_2)}{(\Box_1-\Box_2)}.
\end{eqnarray}
Furthermore, upon the replacement
$g_{\mu\nu}\rightarrow \overline{g}_{\mu\nu}=\phi^2g_{\mu\nu}$
in the quadratic terms of (30), the contributions proportional to
\begin{equation}
\frac{\ln(\Box_1/\Box_2)}{(\Box_1-\Box_2)}
\end{equation}
in (31)--(35) get {\em exactly absorbed} by the second-order
form factors
\begin{equation}
\ln(-\Box/\mu^2) \rightarrow
\ln(-\overline{\vphantom{I}\Box}/\mu^2).
\end{equation}
The remaining tree contributions in (31)--(35)
give rise to the trace anomaly with accuracy
${\rm O}[\Re^3]$ [5], and, as one can check, they are the
expansion terms of the exact action in (24). Thus expression (24)
is in complete agreement with the results of explicit calculations.

The advantages provided by the transition to the
$C^\alpha_{\cdot\beta\gamma\sigma},R$
basis and by the partial summation accomplished in (24) have
an immediate impact on applications. The complicated check [9]
that the $1/\Box,\Box\rightarrow0$, asymptotic terms of the
form factors cancel in the energy-momentum tensor is no more
needed since this cancellation becomes manifest. The contribution
to the energy of the vacuum gravitational waves that was caused
by the $1/\Box$ asymptotic behaviours [8,9] is now generated by a
different, much simpler mechanism. Following the lines of ref. [9],
one must now introduce a new generalized current,
$I^{\mu\alpha\nu\beta}(-\Box,x)$, conjugate to the variation
of the Weyl tensor
\begin{equation}
\delta W = \frac{1}{2(4\pi)^2}\int dx\,g^{1/2}\,
I^{\mu\alpha\nu\beta}(-\Box,x) \delta
C_{\mu\alpha\nu\beta},
\end{equation}
and it turns out that the ${\rm O}(\Box^0)$ term of
$I^{\mu\alpha\nu\beta}(-\Box,x)$
gives directly the quantity
$K^{\mu\alpha\nu\beta}(x)$ which in refs. [8,9]
determined the Bondi--Sachs news functions
${\cal C}_1,{\cal C}_2$ of the vacuum gravitational waves
\begin{equation}
\frac{\partial}{\partial u}
({\cal C}_1+i{\cal C}_2) =
\frac{\partial}{\partial u}
\left\{\frac{r}{8\pi}m_\mu m_\nu
\frac{1}{\Box_{\rm ret}}
\nabla_\alpha\nabla_\beta
K^{[\mu\alpha][\nu\beta]}(x)\right\}\Bigg|_{{\cal I}^+}.
\end{equation}
Here $u$ is the retarded time, $m$ is the complex vector of the
null tetrad, and $r\rightarrow\infty$ is the luminosity distance
to the future null infinity ${\cal I}^+$.

To present an example, we pick out the sector of the effective
action linear in the potential $\hat P$. The
$1/\Box, \Box\rightarrow0$, behaviours of the form factors in
this sector, present in the expansion over the $R^\mu_\nu$
basis [7,9], are now absent completely. On the other hand,
for the contribution linear in $\hat P$ to the vacuum news
functions (39) we obtain straight away
\begin{equation}
K^{\mu\alpha\nu\beta} = \frac{1}{3} {\rm tr}
\frac{1}{(\Box_2-\Box_3)^2}
\left(3\Box_2-\Box_3-2{\Box_2}^2
\frac{\ln(\Box_2/\Box_3)}{(\Box_2-\Box_3)}\right)
C_2^{\mu\alpha\nu\beta}\hat{P}_3.
\end{equation}
The latter expression agrees with the result in [9]
and is of a simpler form.

Finally, one more advantage of the
$C^\alpha_{\cdot\beta\gamma\sigma},R$
basis arises from the fact that, in lorentzian spacetime, the
use of relation (8) imposes no limitation on the initial
state. There is only one solution of eq. (7)  for
$R_{\beta\delta}$ with the appropriate decrease at
${\cal I}^{\pm}$, and this solution is generated by the
retarded Green function for any choice of the in-state.
In contrast with this case, the use of the retarded Green function
in the solution of the Bianchi identities for the Riemann
tensor [1-3] excludes an incoming gravitational wave. This
concerns also eq. (10) above. Normally, the effective action
calculated for a given vacuum remains valid for all coherent
states based on this vacuum. However, because in the calculations
of refs. [3-7] the Riemann tensor was excluded via the Bianchi
identities, the results of these calculations are valid only for
the vacuum state.
The effective action for coherent graviton in-states cannot be
expanded over the $R^\mu_\nu$ basis at all but the
$C^\alpha_{\cdot\beta\gamma\sigma},R$
basis is suitable for this purpose since the only
relation involved is eq. (8).

The present work was supported in part by the Russian
Foundation for Fundamental Research Grant 93-02-15594,
INTAS Grant 93-0493, and the joint Grant MQY300 of the
International Science Foundation and Russian Government.


\begin{thebibliography}{99}
\bibitem{1}
G. A. Vilkovisky,
Class. Quantum Grav. 9 (1992) 895;
preprint CERN-TH.6392/92;
Publ. Inst. Rech. Math. Avanc\'ee, R.C.P. 25, Vol. 43
(Strasbourg, 1992) p. 203.
\bibitem{2}
A. O. Barvinsky, Yu. V. Gusev, G. A. Vilkovisky and V. V. Zhytnikov,
J. Math. Phys. 35 (1994) 3525.
\bibitem{3}
A. O. Barvinsky and G. A. Vilkovisky,
Nucl. Phys. B282 (1987) 163; B333 (1990) 471, 512.
\bibitem{4}
A. O. Barvinsky, Yu. V. Gusev, G. A. Vilkovisky and V. V. Zhytnikov,
J. Math. Phys. 35 (1994) 3543.
\bibitem{5}
A. O. Barvinsky, Yu. V. Gusev, G. A. Vilkovisky and V. V. Zhytnikov,
Nucl. Phys. B439 (1995) 561.
\bibitem{6}
A. O. Barvinsky, Yu. V. Gusev, V. V. Zhytnikov and G. A. Vilkovisky,
Covariant perturbation theory (IV). Third order in the curvature,
University of Manitoba at Winnipeg Report (1993) pp. 1--192.
\bibitem{7}
A. O. Barvinsky, Yu. V. Gusev, V. V. Zhytnikov and G. A. Vilkovisky,
Class. Quantum Grav. 12 (1995) 2157.
\bibitem{8}
A. G. Mirzabekian and G. A. Vilkovisky,
Gravitational waves generated by the vacuum stress,
Phys. Rev. Lett. (1995) to appear.
\bibitem{9}
A. G. Mirzabekian and G. A. Vilkovisky,
Class. Quantum Grav. 12 (1995) 2173.
\bibitem{10}
A. G. Mirzabekian and G. A. Vilkovisky,
Phys. Lett. B 317 (1993) 517.
\bibitem{11}
A. G. Mirzabekian,
Zh. Eksp. Teor. Fiz. 106 (1994) 5
[JETP 79 (1994) 1].
\bibitem{12}
E. S. Fradkin and G. A. Vilkovisky,
Phys. Lett. B 73 (1978) 209.
\bibitem{13}
G. A. Vilkovisky, {\em in} Quantum theory of gravity,
ed. S. M. Christensen (Hilger, Bristol, 1984).
\bibitem{14}
A. O. Barvinsky, A. G. Mirzabekian and V. V. Zhytnikov,
Proc. 6th Quantum Gravity Seminar, Moscow 1995,
to be published.
\bibitem{15}
A. O. Barvinsky and G. A. Vilkovisky,
Phys. Rep. 119 (1985) 1.
\end{thebibliography}
\end{document}